# Understanding Usability and User Acceptance of Usage-Based Insurance from Users' View*

Juan Quintero[1,2]
[1]Uniscon GmbH
Munich, Germany
juan.quintero@fau.de

Zinaida Benenson
[2]Friedrich-Alexander-Universität (FAU)
Erlangen-Nürnberg, Germany
zinaida.benenson@fau.de

## ABSTRACT
Intelligent Transportation Systems (ITS) cover a variety of services related to topics such as traffic control and safe driving, among others. In the context of car insurance, a recent application for ITS is known as *Usage-Based Insurance (UBI)*. UBI refers to car insurance policies that enable insurance companies to collect individual driving data using a telematics device. Collected data is analysed and used to offer individual discounts based on driving behaviour and to provide feedback on driving performance. Although there are plenty of advertising materials about the benefits of UBI, the user acceptance and the usability of UBI systems have not received research attention so far. To this end, we conduct two user studies: semi-structured interviews with UBI users and a qualitative analysis of 186 customer inquiries from a web forum of a German insurance company. We find that under certain circumstances, UBI provokes dangerous driving behaviour. These situations could be mitigated by making UBI transparent and the feedback customisable by drivers. Moreover, the country driving conditions, the policy conditions, and the perceived driving style influence UBI acceptance.

## 1. INTRODUCTION
Automatic tolling, cooperative driving, and traffic control are systems based on communications designed for Intelligent Transportation Systems (ITS) applications [2]. A further innovation in ITS is *Usage-Based Insurance (UBI)*, which is a new trend in the car insurance business. Whereas the traditional car insurance models calculate the premium based on static data (e.g., age, gender, address, car color) and the driving history, UBI calculates premiums based on individual driving style using actual driving data [14]. The traditional car insurance models implement a subsidised system, where better drivers subsidise drivers who have higher accident risk. In contrast, UBI allows fair and personalised policies. The main benefit of UBI for insurers is reduced losses due to more accurate risk calculation [6, 14].

Drivers are also expected to benefit, because UBI insurances incentivise them to improve their driving style through feedback [14]. Thus, UBI has the potential to benefit society due to reduction in traffic congestion, facility costs and the amount of accidents [11].

The main disadvantages of UBI are the investment costs [3] and user privacy [3, 6, 14]. For example, users would not want to disclose some information (i.e., where, when, and how they drive) to insurers, government agencies, or other companies [14]. Usability is an important aspect of UBI solutions, where user interaction with cars and with additional devices is often required. Therefore, people may reject a UBI solution with a low usability. Solutions with high user acceptance and usability have the potential to benefit society. The usability and the user acceptance of current real-world UBI systems are not known. As the first step towards identifying them, we conducted two user studies: a series of semi-structured interviews with UBI users and a qualitative analysis of customer posts about a UBI system called *BonusDrive*. BonusDrive is a UBI program implemented by Allianz (Germany) for young people (18-28 years old) or a family with a young member. In BonusDrive, driving data (i.e., braking, cornering, acceleration, speeding, time of day, and type of road driven) are collected using a telematics device to calculate the driving score. A Bluetooth connection between the telematics device and the car is used for precise vehicle identification.

In this work we identify consequences of using UBI (i.e., dangerous driving, drivers' habit restriction, drivers' privacy decrease), corroborating them by real-world UBI inquiries from an insurer forum. Furthermore, we find that making UBI programs transparent and the feedback customisable by drivers might help to mitigate these consequences. Finally, we identify user acceptance factors related to drivers' external (country driving conditions, policy conditions, and premium reduction) and internal (perceived driving style and perception of UBI) characteristics.

## 2. BACKGROUND AND RELATED WORK
**Usage-Based Insurance** is depicted in Figure 1, where a driver provides her collected driving data to a service provider via a telematics device, such as a Bluetooth dongle (a plug-and-play device), a black box (a professionally installed device), a smartphone or a built-in embedded system that is already present in the recently made cars. Pay-As-You-Drive (PAYD) is often used as a synonym for UBI.

The service provider calculates the driving score and statistics of the driver and sends them to the insurer to calculate the premium discount. The service provider may actually be the same entity as the insurer, but in practice it is often a different company. In this case, the insurer has no access to the behavioral data, as it is processed by a service provider, mitigating users' privacy concerns.

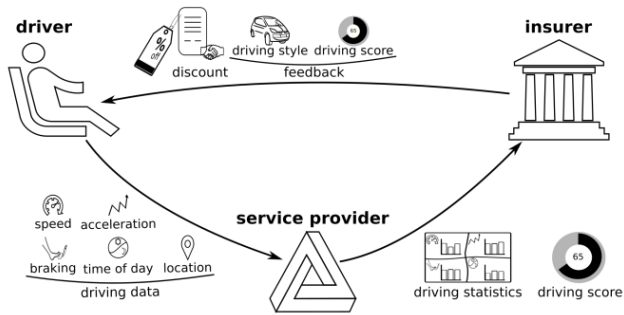

**Figure 1. General Usage-based Insurance model**

**Benefits, Acceptance, and Academic UBI Systems**
Litman [11] compares several distance-based insurance solutions and evaluates potential concerns and benefits for drivers, insurers, and society. Soleymanian et al. [14] find that the drivers improve their driving behaviour after using UBI for six months, decreasing the daily average braking. Mayer [12] studies the acceptance factors for UBI. He finds that the saving potential and the expected effect on driving pleasure are important acceptance factors, but the privacy concerns do not play an important role. Similarly, Derikx et al. [6] find that customers are willing to share their driving data when small financial rewards are offered.

Some academic solutions for UBI have been proposed, but not implemented in practice. Händel et al. [7], Iqbal et al. [8], and Troncoso et al. [15] propose solutions that process driving data locally in the car and send to the insurer only aggregated data to calculate the premium. Boquete et al. [4] propose a similar system with a repository (*Control Centre*) where the insurer can access driving data. Additionally, aggregated data for supporting road vehicle traffic monitoring are provided [7]. Kung proposes a privacy enhancing architecture for ITS called PEARs, which highlights the relevance of architecture in designing a privacy-by-design solution [9].

## 3. METHODOLOGY
We conducted two studies to understand users' attitudes and usage of UBI. A series of semi-structured interviews with UBI users and forum inquiries of a UBI program were analysed using thematic analysis [5] and qualitative content analysis [13], respectively.

### 3.1 Interviews
This study was designed and conducted at an Austrian university during a research internship that originated from a university in Germany. For this reason, we took as reference the UBI programs in Austria and Germany. Originally, we planned to conduct semi-structured interviews with 10 current, 10 former, and 10 potential UBI users, who were defined according to the conditions of UBI programs offered in Austria and Germany. *Current users* are people covered by a UBI program. *Former users* are people who had been covered by a UBI program and for some reason are no longer covered by it. *Potential users* are people over 18 years old with a driving license (never enrolled in a UBI program), who have less than five years of driving experience or are under 30 years old. However, due to recruiting difficulties this plan was later revised (see below). Interviewees were asked about their experiences, opinions, and comments regarding UBI. For potential users we prepared a short video to explain to them what UBI is, without mentioning any advantages or disadvantages of this kind of insurance.

To prepare and to conduct the interviews, first, we developed the interview questions based on academic papers, news, and brainstorming sessions. Second, we submitted the interview questions and the ethical and data protection considerations to the ethics committee at the Austrian university, and received the ethical approval. Third, we recruited the participants and conducted the interviews. Finally, we transcribed the interviews and analysed them using thematic analysis [5].

**Participant Recruitment** Users were recruited via an e-mail list for user studies at the Austrian university, which offers a lot of international programs. We defined the inclusion criteria according to the definition of *current*, *former*, and *potential UBI users*. After running the recruitment questionnaire for five weeks, we got 190 full responses from 3 current, 2 former, and 185 potential users. Therefore, the initial interview plan had to be adjusted. Three current, two former, and twenty (10 female, 10 male) potential users were invited via e-mail to be interviewed, but only one former user and 17 potential users responded to our invitation. Interviewees were compensated with 20€ in cash.

**Data Collection** We conducted the interviews in English, in person, and by Skype. The participants provided an informed consent about data usage and processing. Interviews lasted between 15 and 54 minutes. For each interview we made audio recordings and notes. After conducting interviews, all recordings were transcribed (approx. 586 minutes) and the collected information from interviews (e.g., recordings, notes) was stored with restricted access.

**Data Analysis** We conducted an inductive thematic analysis [5] going from codes to themes. Firstly, one researcher read the transcriptions of interviews and wrote down initial ideas. Secondly, he proposed initial codes for relevant characteristics of the data. Then, he grouped the codes into categories (subthemes) based on the context and the way in which the participants mentioned the codes during the interviews. The themes emerged by grouping similar subthemes together. Finally, the authors discussed the themes and the subthemes, checking the connection between them and the codes. Visualisation tools were used during this process, following the suggestions by Braun and Clarke [5].

### 3.2 Forum Analysis
Allianz forum [1] is an online public space, where the customers can submit inquiries about Allianz products, specific concerns or general questions of insurance. According to the Terms of Use, the users cannot include personal or external data in the posts such as name, address, license plates, among others. Allianz also indicates the users agree to be found in a search for submitted profile or posts information.

**Data Extraction** We selected words related to UBI based on BonusDrive documentation, news, and academic papers. Those words were used as filters (keywords) to extract the data. We implemented a scrape process (i.e., site map, selectors, regular expression including all keywords, and selector graph) using Web Scraper. We performed this process to extract the posts related to BonusDrive and obtained 186 posts.

**Data Analysis** We conducted a qualitative content analysis [13] to analyse the extracted posts. Initially, two researchers independently worked through the first 56 posts to identify relevant categories and to code the posts into identified categories.

Then, they discussed their identified categories, designing a unified code book. Using it, the researchers coded all posts. After all, they obtained a Cohen's Kappa coefficient of greater than 0.80 (an almost perfect agreement according to Landis et al. [10]).

## 4. INTERVIEW FINDINGS

In the interviews with one former and 17 potential UBI users, 3 themes, 15 subthemes and 37 codes were identified, which are presented in Table 1 grouped by themes. Below we describe the identified themes, subthemes, and codes (in *italics*), using the acronym FU (Former User) and PU (Potential User) for interviews quotes.

### 4.1 Advantages

Users mentioned consequences or benefits of using UBI for different actors (i.e., drivers, insurers, and society).

**Benefits for drivers** exhibit that *UBI is a fair system* which provides benefits not only to the insurers, but the drivers such as: *drivers profiling is performed* (i.e., good and bad/reckless), *driving data are connected with each other* (i.e., more accurate estimation of drivers' driving style), and *drivers get additional services* (i.e., emergency call and car location). Other benefits referred were that *drivers identify driving mistakes* through feedback, improving their driving style (*drivers improve driving style*) for getting a premium discount.

**Benefit for insurers** as *insurers' losses are prevented* was reported. Insurers may have less loss if their customers decide to move to UBI, because in UBI insurers could make a more accurate estimation of the drivers' risk based on real driving data.

**Benefits for society** were referred, such as *car accidents decrease*. They have a positive effect on road safety (*security in traffic increases*). PU-12 noted: "if you drive in a safe way you decrease the risk for the insurer and you decrease the overall risk in traffic".

### 4.2 Disadvantages

Users also described some disadvantages of UBI usage.

**Dangerous driving** could be provoked by UBI drivers trying to get a high score or to change their own driving style, causing dangerous situations or accidents (*UBI provokes dangerous driving*). For example, PU-13 noted: "[...] if they (some drivers) receive a score less than 100, they would try another style every next trip. So, they will change their driving style every day and I don't think that it would be very helpful [...] you won't be an attentive and it causes accidents".

**Drivers' habit restriction** is the feeling of many users, who believe that UBI imposes limitations on their driving (*driving habits are restricted*). Thus, UBI pushes drivers to think and to work on their driving before and during a trip. PU-10 claimed: "I don't like to have some limitation when I'm driving, I just like to drive. It's like to have limitations on my driving". Those limitations may provoke that *driving enjoyment decreases*.

**Drivers' privacy decreases** when using UBI (*UBI decreases drivers' privacy*). Although users stated that *people have nothing to hide from insurers*, some users feel uncomfortable knowing that they are being tracked (*drivers feel uncomfortable being tracked*). FU-01 mentioned: "I would not like to be monitored this much". However, other users argued that nowadays other popular technologies are not privacy-respecting (*other technologies do not provide privacy*) and UBI could be one of those.

**Table 1. Thematic analysis from interviews with users**

| Theme | Subtheme | Code |
|---|---|---|
| Advantages | Benefits for drivers | Driving data are connected with each other |
| | | Drivers get additional services |
| | | Drivers identify driving mistakes |
| | | Drivers improve driving style |
| | | Drivers profiling is performed |
| | | UBI is a fair system |
| | Benefit for insurers | Insurers' losses are prevented |
| | Benefits for society | Car accidents decrease |
| | | Security in traffic increases |
| Disadvantages | Dangerous driving | UBI provokes dangerous driving |
| | Drivers' habit restriction | Driving enjoyment decreases |
| | | Driving habits are restricted |
| | Drivers' privacy decreases | Drivers feel uncomfortable being tracked |
| | | Other technologies do not provide privacy |
| | | People have nothing to hide from insurers |
| | | UBI decreases drivers' privacy |
| | Feedback not customisable | Feedback during trip distracts drivers |
| | | Feedback should be customisable by drivers |
| | Scoring system | Driving data do not determine driving skills |
| | | Driving data may be wrong |
| | Telematics device | Drivers have to configure the telematics device before each trip |
| | | Telematics device may be outdated |
| | Use of driving data for other purposes | Someone may use driving data for other purposes |
| Acceptance Factors | Country driving conditions | Driving traditions influence drivers' habits |
| | | Relaxed regulation prevents from improving driving skills |
| | Perception of UBI | Insurer's reputation |
| | | Negative experience of UBI users |
| | | People do not trust insurers |
| | | UBI opinion of a relative, friend, or expert |
| | | Service provider's reputation |
| | Policy conditions | Insurance conditions |
| | | Insurance costs |
| | | UBI transparency |
| | Premium reduction | Insurance discount |
| | | Premium discount motivates to change driving style |
| | Perceived driving style | Monitored drivers drive better |
| | | Own driving style perception |

**Feedback not customisable** was reported. Users suggested that *feedback should be customisable by drivers*. They mentioned that getting feedback or advice during a trip could be distracting and annoying for the drivers (*feedback during trip distracts drivers*). In this situation the drivers could pay more attention to the feedback instead of the road.

**Scoring system** of UBI is based on the collected driving data, which allows that the drivers get a score. Users criticised the collected driving data, stating that the "time of the day" and "GPS location" cannot be adjusted without changing the driver's lifestyle. Furthermore, they argued that the driving data are not enough to determine driving style (*driving data do not determine driving skills*). Therefore, UBI should consider other external factors (e.g., pedestrians, traffic conditions, weather, driving history) to get an accurate driving score and driving behaviour. Some users argued that the telematics device may gather wrong (not accurate) driving data due to GPS signal loss or some technical problems (*driving data may be wrong*), which affect their driving score.

**Telematics device** has to be configured by drivers before each trip (*drivers have to configure the telematics device before each trip*) to collect the driving data. That is a hassle for some users, especially FU-01 claimed: "[...] you get in the car, you're got to go like! I forgot to put it ON again. Then, you have to stop, to put it ON and you can leave". Additionally, some users were concerned that the *telematics device may be outdated*.

**Use of driving data for other purposes** was reported as the main disadvantage of using UBI. Users mentioned that someone (i.e., insurer, service provider, other people) could use their driving data without authorization for other purposes (*someone may use driving data for other purposes*) such as customer profiling, marketing, solving accident investigations, or identifying moving traffic violations (e.g., driving over speed, breaking signals, street racing, etc.).

### 4.3 Acceptance Factors
We identified important factors for users to make a decision about being covered or not by UBI. These factors are described below.

**Country driving conditions** are referred as rules (e.g., speed limit, parking), regulations (e.g., laws, data protection), and drivers' driving behaviours in a specific country where UBI is offered. Some users consider it too complicated to have a good driving style due to the influence of country driving conditions on drivers' habits (*driving traditions influence drivers' habits*). Also, they referred *relaxed regulation prevents from improving driving skills* such as: PU-10 "you go to Greece and then you see people from England, Germany, and other strict countries how they behave, why they're not acting like they are acting in England or Germany? They have police stricter than in Greece".

**Perception of UBI** is understood as the *UBI opinion of a relative, friend, or expert* which is important to the users at the moment to make a decision about UBI. This opinion is based on information about *insurer's reputation*, *service provider's reputation*, and users' reviews. Some users highlighted negative experiences as the most relevant in perception of UBI. Thus, *negative experience of UBI users* (related to the insurer, the service provider, the UBI program, or the telematics device) provides users insight about potential issues. Usually users need more information about UBI because *people do not trust insurers*.

**Policy conditions** such as *insurance conditions* and *insurance costs* were mentioned. Users remarked program conditions (e.g., coverage, kilometers driven, discount) and contract conditions (e.g., term of use, cancellation conditions) as important information for making a decision about UBI. In addition, some users want to know more details about *UBI transparency* (e.g., who can access their driving data, where are their data stored, when will be their data deleted, among others).

**Premium reduction** was reported as the main users' goal in UBI. Users want to save money reducing their premiums, finding in UBI a way to do it (*insurance discount*). FU-01 mentioned: "the main reason would be the money because ensuring your car is never cheap". Thus, users found that a *premium discount motivates to change driving style*, such as: PU-06 "the discount, I guess you have like a motivation to drive safely and correctly".

**Perceived driving style** is defined as how the users perceive their own driving style. According to that, they could decide to join (good driving style) or not (bad driving style) to a UBI program. Users described "good driving style" as following the rules, being careful and respectful with other drivers (*own driving style perception*). Users mentioned that usually drivers adapt their driving behaviour when they know that someone is monitoring them (*monitored drivers drive better*).

## 5. FORUM FINDINGS
In the qualitative content analysis of Allianz forum, 12 categories were identified. These categories are presented in Table 2, where the topics that corroborate our findings from interviews with one former and 17 potential UBI users are labeled with **\***. We grouped the categories into four general classes, which are described below. Categories are presented in *italics* and the acronym CU (Current User) is used for presenting forum quotes.

**Table 2. Categories from Allianz forum (186 posts), topics labeled with \* were also discussed in the interviews. The number of posts of each category is indicated in brackets ( ). Some posts included more than one category.**

| Category | Explanation |
|---|---|
| *Wrong score (39) | Score obtained for a trip is different from the driver's expectation |
| Trip recorded, but not saved (31) | App showed ongoing recording, but trip was not shown in the logbook afterwards |
| App stopped working (25) | After updating the smartphone or changing app properties the app stops working |
| Starting problems (25) | Due to logistical or technical problems the user cannot begin to use BonusDrive |
| Trip not recorded (21) | Trip recording process did not start or trip was only partially recorded |
| *Problems with Bluetooth (16) | Bluetooth pairing process does not work, or connection breaks down |
| Distrust in the insurer (15) | User suspects Allianz provides wrong, unrealistic or contradictory information on purpose |
| *BonusDrive provokes dangerous driving (11) | Trying to improve driving score, users generate dangerous situations |
| *Trips are recorded although users do not drive their cars (10) | Recording trips in passenger mode or when the user travels by foot, train, bike or in another car |
| *Trip recording starts too late (9) | Trip recording process starts later than the actual trip, leading to wrong scores |
| Information is lost (6) | Users cannot see previously saved information in the app |
| App takes too long to save the trip (3) | Users have to wait in the car several minutes after the trip end till the trip is saved |

**Scoring system** Many users indicated that they got a *Wrong score*. They criticised the criteria to evaluate the driving behaviour. Some users reported logistical or technical *starting problems* that prevent them from getting a score, especially often they did not receive in time the Bluetooth dongle needed for trip recording. In addition, sometimes *trips are recorded although*

*users do not drive their cars*, such that data from train, bike, or another car were collected.

**Driving data loss** User reported losing their collected driving data (*trip recorded, but not saved*). Some users claimed that the trips are not recorded at all or only partially (*trip not recorded*). In some cases previously recorded *information is lost*, meaning that saved driving data cannot be seen in the app. Thus, the users cannot get any score.

**Telematics device issues** The users referred that sometimes *app stopped working*. Other times, when the user does not wait for a few minutes after the trip end, the app shows an error and no data of this journey are saved (*app takes too long to save the trip*). Furthermore, *trip recording starts too late* was also reported due to *problems with Bluetooth* or GPS connection problems.

**Consequences of using UBI** Users expressed *distrust in the insurer*, assuming the insurer might affect their driving score on purpose. Furthermore, they argued that trying to improve their driving score, they could generate dangerous situations or cause an accident (*BonusDrive provokes dangerous driving*). CU-107 claimed: "I have myself already provoked the anger of other drivers because I drove 20 km/h on a federal highway with curves just to get a good rating. But, this also did not happen!"

Most of the identified categories described above are related to usability issues. Users were concerned about getting a good driving score. During this process, they identified different situations which represent a barrier to achieve their goals. Moreover, some of these situations were anticipated in the interviews by potential users, and also mentioned by the former user.

## 6. DISCUSSION
In this section, we discussed consequences of using UBI and two identified acceptance factors.

### 6.1 Consequences of UBI usage
Users criticised that UBI may reduce the enjoyment of driving and the drivers' privacy, as well as provokes dangerous driving. They argued that to get a good driving score in UBI, drivers try to adapt their driving style according to UBI criteria (usually unknown). Thus, drivers change their driving habits, making the driving process into a monotonous activity without any enjoyment. In UBI the drivers generally get feedback or advice during a trip, which could be distracting and annoying for them. Therefore, drivers generate potential risks to road traffic trying to follow the feedback from a non-transparent system to get a higher driving score. That concern was corroborated in the forum (BonusDrive provokes dangerous driving) by users in a real-world UBI program.

Regarding drivers' privacy, users realised that from GPS location could be inferred patterns of behaviour, which someone may use not only for calculating the premium or providing feedback.

### 6.2 Acceptance factors
Users consider that country driving conditions influence on their driving performances. Thus, they could be not motivated to join to UBI in a country where the drivers have bad driving habits and the regulations are not very strict. Moreover, only people who consider themselves as "good drivers" are willing to join to UBI (perceived driving style) for getting a premium reduction. The process to identify themselves as a "good or bad" driver is very subjective taking into account that people usually do not recognise their mistakes. So, after joining to UBI, users who consider themselves as a "good" driver could leave UBI if they will not get a discount, even though they improve their driving style.

### 6.3 Limitations
Although we got valuable insights from our interviews with UBI users, the number of interviewees was limited, maybe due to the few UBI programs offered in Europe. We also considered an online forum of a single UBI solution in Germany, which does not allow us to generalise our findings.

## 7. CONCLUSION AND NEXT STEPS
We analysed and identified consequences of using UBI, corroborating them by an analysis of usage problems of a real-world UBI system. We conclude that drivers could generate dangerous situations trying to improve their driving style with non-transparent UBI programs, representing threats to road traffic safety. Additionally, we found that driving traditions and regulations of the country where UBI is offered, as well as users' perception of their own driving style and the policy conditions influence UBI acceptance. We recommend insurers to make UBI transparent (e.g., data usage, evaluation criteria) and the feedback customisable by drivers according to their preferences. Thus, the identified consequences of using UBI could be mitigated.

Our next step for this project is to build a user acceptance model using the findings of this study. This model will be validated via an online survey.

## 8. ACKNOWLEDGMENTS

This research was supported by the "Privacy&Us" Innovative Training Network (EU H2020 MSCA ITN, GA No.675730). We would like to thank Alexandr Railean for useful comments and discussions.

## 9. REFERENCES

[1] Allianz Kundenservice. 2019 (accessed May 1, 2019) https://forum.allianz.de/hilfe/kfz-versicherung?page=1

[2] Andrisano, O., Verdone, R., & Nakagawa, M. 2000. Intelligent transportation systems: the role of third generation mobile radio networks. *IEEE Communications Magazine*, 38(9), 144-151.

[3] Arvidsson, S. 2011. *Reducing asymmetric information with usage-based automobile insurance*. Technical Report. Swedish National Road and Transport Research Institute.

[4] Boquete, L., Rodríguez-Ascariz, J. M., Barea, R., Cantos, J., Miguel-Jiménez, J. M., & Ortega, S. 2010. Data acquisition, analysis and transmission platform for a pay-as-you-drive system. *Sensors*, 10(6), 5395-5408.

[5] Braun, V. and Clarke, V. 2006. Using thematic analysis in psychology. *Qualitative research in psychology*, 3(2):77–101.

[6] Derikx, S., de Reuver, M., and Kroesen, M. 2016. Can privacy concerns for insurance of connected cars be compensated? *Electronic Markets*, 26(1):73–81.

[7] Händel, P., Ohlsson, J., Ohlsson, M., Skog, I., & Nygren, E. 2014. Smartphone-based measurement systems for road vehicle traffic monitoring and usage-based insurance. *IEEE systems journal*, 8(4), 1238-1248.

[8] Iqbal, M. U., & Lim, S. 2006. A privacy preserving GPS-based Pay-as-You-Drive insurance scheme. In *Symposium on GPS/GNSS (IGNSS2006)*, pages 17–21.

[9] Kung, A. 2014. Pears: privacy enhancing architectures. In *Annual Privacy Forum*, pages 18–29. Springer, Cham.

[10] Landis, J. R. and Koch, G. G. 1977. The measurement of observer agreement for categorical data. *Biometrics*, 33(1):159–174.

[11] Litman, T. 2011. *Distance-based vehicle insurance feasibility costs and benefits*. Technical Report. Victoria Transport Policy Institute.

[12] Mayer, P. 2012. *Empirical Investigations on User Perception and the Effectiveness of Persuasive Technologies*. Doctoral Thesis. University of St. Gallen.

[13] Saldaña, J. 2009. *First cycle coding methods. The Coding Manual for Qualitative Researchers*. Thousand Oaks, CA: Sage Publications Ltd.

[14] Soleymanian, M., Weinberg, C. B., & Zhu, T. 2019. Sensor Data and Behavioral Tracking: Does Usage-Based Auto Insurance Benefit Drivers? *Marketing Science*, 38(1), 21-43.

[15] Troncoso, C., Danezis, G., Kosta, E., Balasch, J., and Preneel, B. 2011. Pripayd: Privacy-friendly pay-as-you-drive insurance. *IEEE Transactions on Dependable and Secure Computing*, 8(5):742-755.